\documentclass[aps, pre, twocolumn, floatfix, amssymb, amsmath]{revtex4-1}

\usepackage[applemac]{inputenc}

\usepackage[usenames,dvipsnames,svgnames,table]{xcolor}
\usepackage{graphicx}

\usepackage{natbib}%
\usepackage{amsbsy}
\usepackage{gensymb}
\usepackage{CJK} 
\usepackage{psfrag}
\usepackage{siunitx}
\usepackage{color,soul}

\newcommand{\sect}{Sec. }
\newcommand{\fig}{Fig.} 
\newcommand{\ie}{i.e.}
\newcommand{\eg}{e.g.}
\newcommand{\zst}{z_c(r,t^{\infty})} 
\newcommand{\zdyt}{z_c(r,t)} 
\newcommand{\zdy}{z_c(r,t^*)}
\newcommand{\zstxy}{z_c(x,y,t^{\infty})} 
\newcommand{\dcinf}{D_c^{\infty}} 
\newcommand{\dzmax}{D_c^*} 
\newcommand{\zmax}{Z_c^*} 
\newcommand{\cangle}{\theta_c^{\infty}} 
\newcommand{\cangleD}{\theta_c^{*}} 
\newcommand{\abc}[1]{(\textit{#1})}
\newcommand{\We}{\textrm{We}} 

\begin{document}
\begin{CJK*}{UTF8}{}
\title{Crater formation during raindrop impact on sand}
\author{Rianne de Jong}
	\email{jong.riannede@gmail.com}
\author{Song-Chuan Zhao \CJKfamily{gbsn}(赵松川)}
	\email{songchuan.zhao@outlook.com}
\author{Devaraj van der Meer}
\affiliation{Physics of Fluids Group, MESA+ Institute for Nanotechnology, and J.M. Burgers Center for Fluid Dynamics, University of Twente, P.O. Box 217, 7500 AE Enschede, The Netherlands}
\date{\today}

\begin{abstract}
After a raindrop impacts on a granular bed, a crater is formed as both drop and target deform. After an initial, transient, phase in which the maximum crater depth is reached, the crater broadens outwards until a final steady shape is attained. By varying the impact velocity of the drop and the packing density of the bed, we find that avalanches of grains are important in the second phase and hence, affect the final crater shape. In a previous paper, we introduced an estimate of the impact energy going solely into sand deformation and here we show that both the transient and final crater diameter collapse with this quantity for various packing densities. The aspect ratio of the transient crater is however altered by changes in the packing fraction.
\end{abstract}
\maketitle
\end{CJK*}

\section{Introduction}
The impact of a droplet on a bed of sand or soil is amongst the most common events in nature; it is observed in agriculture, where rain is needed for proper crop growth; and it occurs in industry in the process of wet granulation which is the basis of the production of many pharmaceuticals. The rich phenomenon of drop impact on granular matter has only started to draw attention recently \cite{Katsuragi2010, Marston2010, Katsuragi2011, Delon2011, Nefzaoui2012, Emady2013a, Long2014, Zhao2015PNAS, ZhaoSC2015, Zhang2015, Zhao2017} and its underlying physics is still largely unrevealed. One of these aspects that is not yet fully understood, is the formation of impact craters during droplet impact, which is the focus of this work.

In the literature, craters formed by solid intruders have been observed from small laboratory to very large scale geological events, where the development and shape of these (planetary) craters continuously intrigued the scientific field for more than thirty years \cite{Pike1980, Melosh1989, Holsapple1993, Uehara2003, Walsh2003, Lohse2004PRL, deVet2007, Marston2012a, Ruiz-Suarez2013}. \citet{Melosh1989} divided the evolution of a geological crater into three phases: First, the initial contact, almost instantly followed by a shock wave through the granular bed. Second, the transient crater growth and third, slumping of the crater walls and/or collapse of the splash (or jet) at the center \cite{Melosh1989, Ruiz-Suarez2013}. These three phases have also been observed in laboratory experiments for solid intruder impact on dry grains \cite{Uehara2003, Walsh2003, Lohse2004PRL, Marston2012a, Clark2015}. If impact energy is lowered, if gravitational energy is increased or if the target is harder to permeate, the third phase may not occur \cite{Holsapple1993} and, for dry grains, replaced by avalanches.

Impact of droplets on grains results in a very different type of complex behavior. Observations for single drop impact on (dry) grains, show that both intruder and target deform, and mixing occurs (for wettable grains) \cite{Katsuragi2010, Marston2010, Katsuragi2011, Delon2011, Nefzaoui2012, Emady2013a, Long2014, Zhao2015PNAS, ZhaoSC2015, Zhao2017}, and recently an analogy between the raindrop imprints and asteroid strikes was drawn~\cite{Zhao2015PNAS}. In contrast to solid ball impact on sand, droplet deformation also consumes part of the kinetic energy $E_k$ of the impactor, and the energy partition depends on the stiffness of both intruder and target \cite{ZhaoSC2015, Zhao2017}. We will show that the deformability of the intruder can significantly affect crater formation. This is consistent with what is found for deformable granular projectiles impacts on sand \cite{Pacheco2011}.

The sand bed response after drop impact is mainly investigated by means of determining the final crater diameter, for which various scaling laws are obtained \cite{Katsuragi2010, Katsuragi2011, Delon2011, Nefzaoui2012, Zhao2015PNAS}. The maximum depth of the crater usually cannot be measured from the craters final profile as the irregularly shaped liquid-grain residue obstructs it, but it may be obtained in the cases where the droplet (or a mixture of droplet and grains) rebounces \cite{Zhao2015PNAS}. In \cite{ZhaoSC2015}, however, high-speed laser profilometry during impact is used to acquire the transient behavior of the crater depth. 

In this work we study both the transient and final crater shape, focusing on the evolution from the moment in time $t^*$ at which the maximum crater depth is reached until the crater obtains a final steady shape (\ie, at some point in time $t^{\infty}$ long after the impact). To our knowledge, both for solid and deformable intruders impacting on sand, this comparison has not been made. 
In particular, we will show that avalanches play a key role in the post-impact evolution of the crater, which is a mechanism that is not traditionally accounted for in the literature. In addition, we study the effect of both the initial drop energy $E_k$ and the packing density of the bed. The structure of the paper is as follows: We will first discuss the experimental method (\sect \ref{sec:method}), after which we start the result section, \sect \ref{sec:results}, with discussing the final crater shape and how it depends on the impact parameters. It is followed by the crater evolution (\sect \ref{sec:evolution}), from the perspectives of the slope of the crater and the displaced volume. Finally, we review and compare our results with previous works in \sect \ref{sec:discussion}, where various scaling laws will be discussed in more detail.

\section{Experimental method \label{sec:method}}
\begin{figure}\begin{center}
	\includegraphics[width=8.6cm]{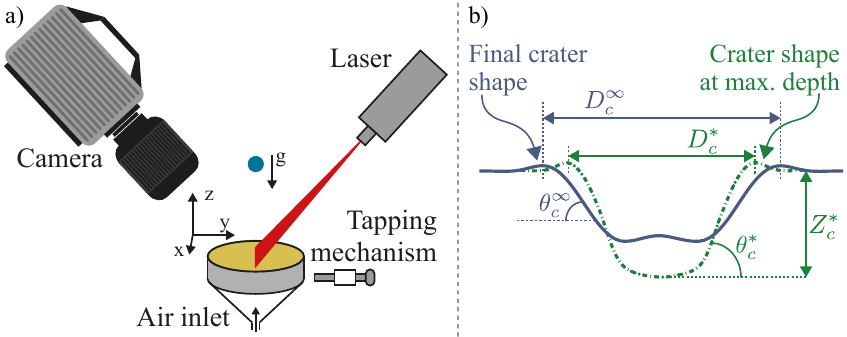}
	\caption{\abc{a} Experimental setup of droplet impact on a granular bed. The height profile, $z(x,y,t)$ is measured with laser sheets and high speed camera. \abc{b} Relevant length scales of the crater at the moment it reaches its maximum depth and at its final shape. 
	\label{fig:setup} }
\end{center}
\end{figure}
\begin{figure}\begin{center}
	\includegraphics[width=8.6cm]{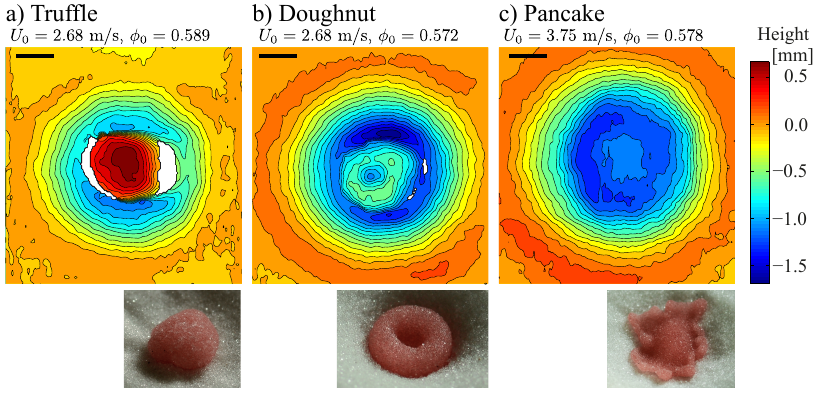} 
	\caption{ Height profiles, $\zstxy$, obtained by scanning the crater after impact. Typical liquid-grain residues are denoted as truffle \abc{a}, doughnut \abc{b} and pancake \abc{c}, and of each an example picture is added, where the liquid was colored red, to increase contrast. The bar at the top left of each figure represents 2 mm.  White areas in the contour plot lack height information as either laser or camera is blocked due to a sharp height change (as the scan is executed with a translation of the substrate only).	
	\label{fig:contours} }
\end{center}
\end{figure}

The intruder, an ultrapure (milli-Q) water droplet (density $\rho_l = 998$~kg/m$^3$, surface tension $\sigma = 73$~mN/m, viscosity $\mu = 1.0$~mPa/s), is released from a needle when gravity overcomes surface tension, resulting in a droplet with a diameter of $D_0 = 2.84\pm 0.03$~mm. The flow rate of the syringe pump is $0.15$~ml/min, which is sufficiently small to minimize inertial effects and obtain a reproducible droplet size. The droplet is released from various heights, giving it an impact speed $U_0$ varying between $1.3-4.2$~m/s, which is determined from a calibrated height-impact speed curve. As target, a granular bed of at least $23$~mm in height and $39$~mm in diameter is used, consisting of polydisperse glass beads with a diameter $d_g = 70-110\  \mu$m and specific density $\rho_g = 2.5 \cdot 10^3$~kg/m$^3$. Before the experiments, the beads are dried in the oven for at least half an hour. The bed is carefully prepared by air fluidization and optionally compacted with a tapping device (\fig~\ref{fig:setup}\textit{a}) to reach a variety of packing fractions between $0.54 < \phi_0 < 0.61$. The latter is calculated by means of weighing the grains in combination with measuring the surface height before impact relative to the container rim (along the line of major height variation, \ie, the direction of tapping).

We study the effect of droplet impact on grains during as well as after the impact. To obtain both horizontal and vertical substrate deformation, we shine a laser sheet (with a depth of focus of $20.4$~mm and width of $79$~$\mu$m) across the granular bed and focus a high speed camera (Photron SA-X2) on the target surface, see \fig~\ref{fig:setup}\textit{a}. Laser and camera angle are kept between $45\si{\degree}$ and $65\si{\degree}$ with the substrate. During impact, two parallel laser sheets are used, in combination with a camera frame rate of $10,000$~fps, to acquire a radial height profile $\zdyt$ by assuming axisymmetry. 
When the location of impact is far from the laser lines, the part of $\zdyt$ close to the crater center ($r=0$ mm) is unresolved. We discard the experiments where the unresolved region is beyond $r=1$ mm, and extrapolate the crater profile to $r=0$ with a hyperbola fit otherwise. For more details, see \cite{ZhaoSC2015}. Some minutes after impact, the final crater shape is scanned by carefully translating the substrate horizontally through the laser sheet (accuracy: $16$~$\mu$m/frame), such that a full height profile $\zstxy$ can be obtained after analysis, see \fig~\ref{fig:contours}. When reconstructing the profile $\zstxy$, we set the horizontal resolution in the scan direction to the smallest grain size ($70$~$\mu$m) \footnote{Note that as the laser is under an angle, we also take into account the horizontal shift when reconstructing the depth $\zstxy$.}. In the vertical direction and in the direction along the laser line the minimal resolution is $0.1$~mm per pixel.

From this three-dimensional crater shape $\zstxy$ we reconstruct a $2$D profile $\zst$, by acquiring the crater center and averaging over the azimuthal coordinate. In this procedure, the presence of a liquid residue at the crater base necessitates careful analysis. Hence, to locate the crater center, we restrict ourselves to the crater walls and only consider height contours which are circular. The centers of these contours are averaged and consequently $\zst$ is obtained.

\section{Results \label{sec:results}}
\subsection{Final crater profile\label{sec:cprofile}}
\begin{figure*}\begin{center}
	\includegraphics[width=1.0\textwidth]{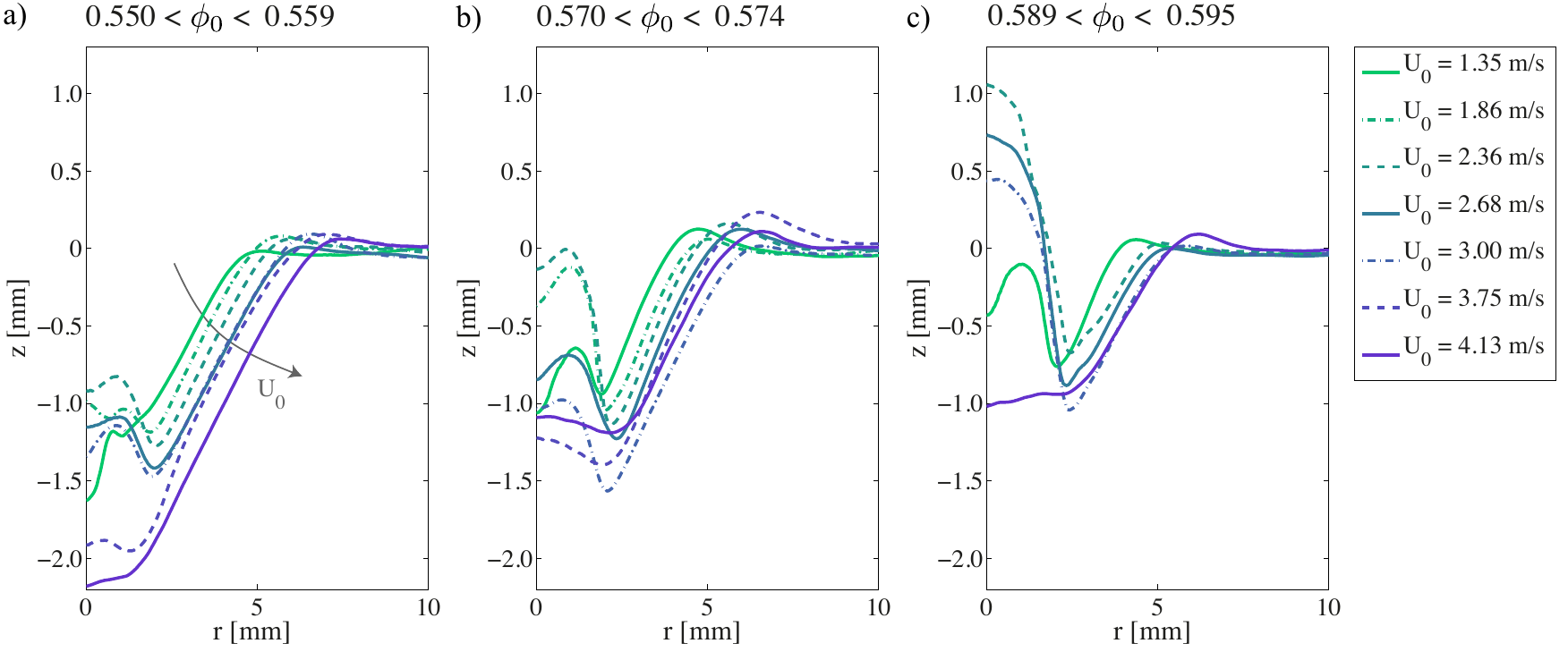}
	\caption{Various radial height profiles $\zst$ of the final crater are shown for increasing impact velocities and, from left to right, for increasing packing fraction. Note that the center region is occupied by the granular residue, which for all packing fractions changes from a doughnut, through a truffle to a pancake shape for increasing velocity.
	\label{fig:zofr} }
\end{center}
\end{figure*}

A selection of these averaged final crater profiles $\zst$ is illustrated in \fig~\ref{fig:zofr}, which shows that the crater shape, diameter and depth strongly, and sometimes even non-monotonously, depend on impact velocity $U_0$ and packing fraction $\phi_0$. One can observe that the final crater depth, measured adjacent to the granular residue, $r \approx 2$mm, decreases for denser packings, which is plausible as a dense sand bed behaves more solid-like and hence, it is more difficult to deform. The maximum crater depth is reached, however, during impact rather than at its final shape, as contraction of the liquid-grain residue and avalanches tend to make the crater shallower. Therefore, in this work we will distinguish two phases: the initial crater opening until its maximum depth is attained and the crater evolution from this moment onward until the final shape has been reached. Besides the above effect on the crater shape, also the shape of the granular residue, \ie, the liquid-grain mixture that remains in the center of the crater after impact, is altered greatly with $U_0$ and $\phi_0$. A theory, developed in \cite{ZhaoSC2015}, explains well the transition observed from the contracted doughnut and truffle residue to the flat pancake shapes (\textit{cf.} \fig~\ref{fig:contours} and \ref{fig:zofr}). Here we will concentrate on the behavior of the characteristic length scales of the crater as a function of velocity and packing fraction, and discuss this in the following (sub)sections for both the final and the transient case.

\subsection{Final crater diameter \label{sec:diam}} 
\begin{figure}\begin{center}
	\includegraphics[width=8.6cm]{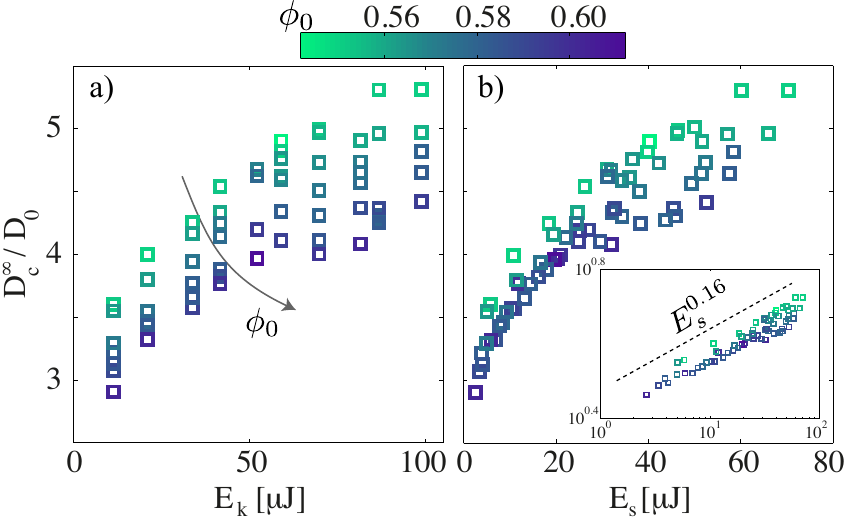}
	\caption{Dimensionless final crater diameter versus (\textit{a}) the kinetic energy $E_k$ of the impacting droplet, (\textit{b}) the estimated amount of kinetic energy transferred to the sand $E_s =  [\zmax / \left(\zmax + D_0/2 \right)] E_k $. The inset shows the same data on a doubly logarithmic scale. The dashed line indicates $\dcinf/D_0 \propto E_s^{0.16}$. } 
	\label{fig:dmax_final} 
\end{center}\end{figure}
The final crater diameter $\dcinf$, which is defined in this work as the diameter of the final crater shape at the maximum height of the rim (see \fig~\ref{fig:setup}\textit{b}), is shown in \fig~\ref{fig:dmax_final}, in which it is non-dimensionalized with the droplet diameter $D_0$.
Figure \ref{fig:dmax_final}\textit{a} displays the data against the kinetic energy of the intruder, $E_k = \frac{\pi}{12} \rho_l D_0^3 U_0^2$. Here it is seen that the diameter $\dcinf$ increases with kinetic energy, but scatters greatly with packing fraction. As both intruder and target deform during impact, part of the initial kinetic energy will be transferred into droplet deformation ($E_d$) and another part into sand deformation energy ($E_s$). In \citep{ZhaoSC2015}, a division of the kinetic energy is proposed related to the stopping force that the drop experiences, $F = E_k / (D_0/2 + \zmax)$ with $\zmax$ the maximum depth of the crater (\fig~\ref{fig:setup}\textit{b}). Multiplying the stopping force $F$ with the penetration depth $\zmax$ leads to the following estimate of the energy that is transferred to the granular bed $E_s = \frac{\zmax}{D_0/2 + \zmax} E_k$. When we plot the final crater diameter data $\dcinf/ D_0$ versus this rescaled energy $E_s$ (\fig~\ref{fig:dmax_final}\textit{b}), we obtain a good collapse of the data for various packing fractions, at least for low energies. As indicated in the inset of \fig~\ref{fig:dmax_final}\textit{b}, the data are consistent with a power-law scaling of $\dcinf/D_0$ with $E_s^{\alpha}$ with $\alpha \approx 0.16 \pm 0.02$, obtained from a weighted robust fit. The data confirm the hypothesis that a denser packing leads to a shallower crater due to a smaller energy adsorption.

For the higher energy data in \fig~\ref{fig:dmax_final}\textit{b}, there is however a significant spread with packing fraction. Various mechanisms may be important: After the crater reaches its maximum depth, the crater evolves further, as grains shoot away \cite{Long2014} and avalanches occur before reaching to a final state in which the crater diameter is maximal. As a consequence, the final crater shape is formed by complex and subtle processes and therefore, we will focus in the rest of the result section on the evolution of the crater from the moment it reaches its maximum depth onto the final crater shape.

\subsection{Crater evolution \label{sec:evolution}}

\begin{figure}\begin{center}
 	\includegraphics[width=8.6cm]{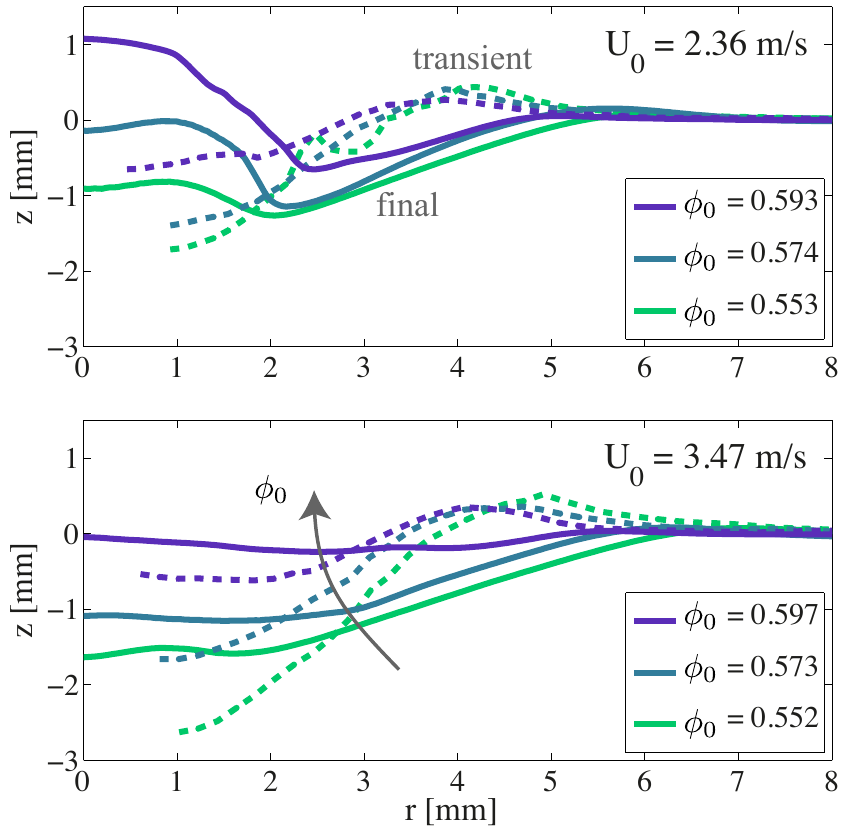}
	\caption{Transient and final radial height profiles $\zdy$~(dashed lines) and $\zst$~(solid lines) for two impact velocities and three different packing fractions. As the droplet impact location is usually at a distance from the laser lines, data for the transient height profile near the center ($r=0$) is missing.
	\label{fig:zofr_sd} }
\end{center}
\end{figure}

In \fig~\ref{fig:zofr_sd}, the transient crater profile at the moment of reaching the maximum depth, $\zdy$ and the final profile $\zst$ are compared. One clearly observes how the crater evolves after reaching its largest depth: The crater rim broadens, its depth decreases and the slope of the crater wall diminishes (both for the doughnut/truffle and the pancake morphology of the residues presented in \fig~\ref{fig:zofr_sd}\textit{a} and \textit{b} respectively). Furthermore, it evidences that the crater shape is strongly affected by the packing density, especially for the high impact velocities.

\begin{figure*}
\begin{center}
 	\includegraphics[width=\textwidth]{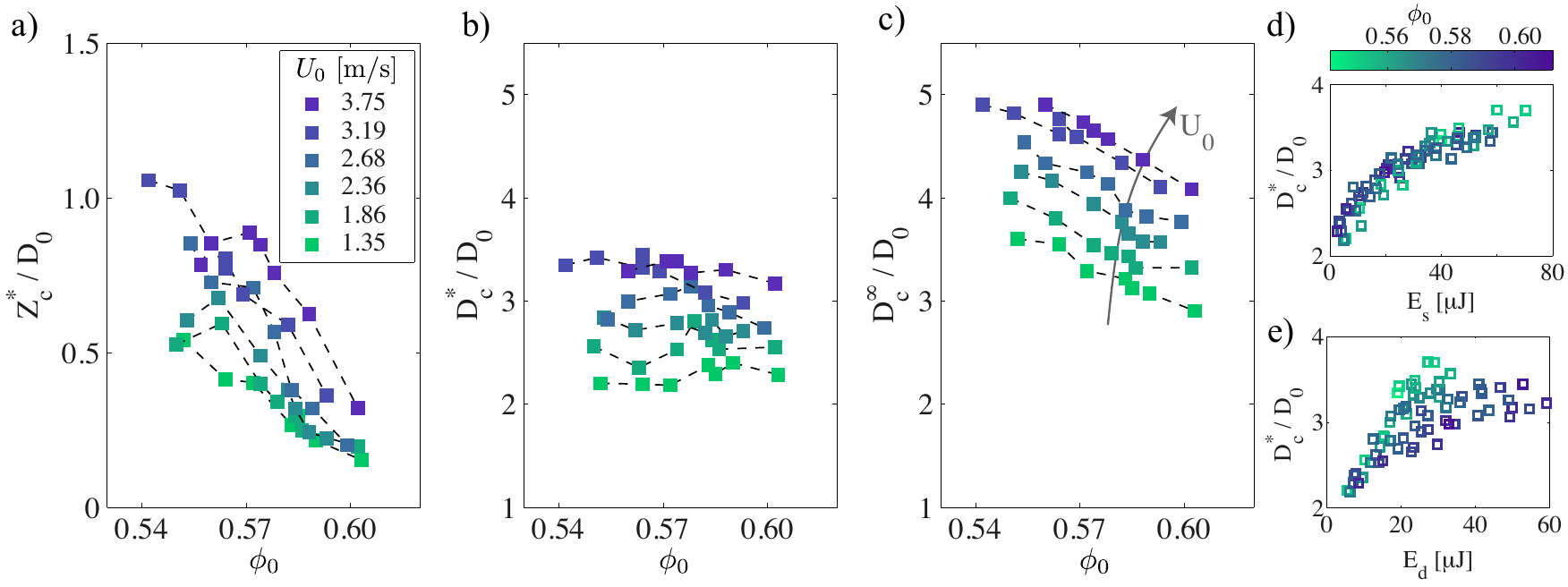}
	\caption{ \abc{a-c} Relevant length scales non-dimensionalized with the initial drop diameter $D_0$ are plotted against packing fraction with in \abc{a} the maximum depth $\zmax$, in \abc{b} the crater diameter $\dzmax$ at the moment the maximum crater depth is reached, and in \abc{c} the maximum diameter $\dcinf$ of the final crater. The legend in \abc{a} corresponds to all  three subfigures. \abc{d-e} The transient diameter $\dzmax / D_0$ is plotted  versus the energy $E_s$ transferred to the sand \abc{d} and versus the energy $E_d$ transferred to the droplet \abc{e}. Clearly $E_s$ does collapse the data whereas $E_d$ does not. For both figures the colorbar at the top indicates the packing fraction.
	\label{fig:maxZDD} }
\end{center}
\end{figure*}
To capture this quantitatively, in \fig~\ref{fig:maxZDD}\textit{a-c} we look at the relevant length scales extracted from these $z(r)$-plots, namely the maximum depth $\zmax$, the diameter at the moment of reaching the maximum depth $\dzmax$ and the final crater diameter $\dcinf$, as defined in \fig~\ref{fig:setup}\textit{b} \footnote{The crater diameter can also be determined at $z=0$, \ie, the excavation diameter \cite{Holsapple1993, Walsh2003}. For the transient case, we can trustfully do so and it consistently gives a result of approximately $0.8 \dzmax$. For the final crater shape, the zero base line cannot be determined without ambiguity, and hence, we don't calculate it.}.
We observe that $\zmax$ and $\dcinf$ noticeably decrease with packing density, but surprisingly, that the transient diameter, $\dzmax$, hardly varies with $\phi_0$. Only for the high velocity cases, one may observe a small dependence. Therefore, the shape and diameter of the crater rim must be affected by the packing fraction mainly in a later stage. This gives the impression that  events occurring after the opening of the crater, such as splashing \cite{Boudet2006, Marston2012a} and avalanches \cite{Bonnet2010, Rondon2011, Gravish2014}, shape the final crater rim diameter in a profound way. Furthermore, \fig~\ref{fig:maxZDD} shows that the transient crater diameter and depth respond differently to packing fraction variations, which we will discuss further in section \ref{sec:depthdiam}.

What could be the initial driving mechanism for the formation of the crater rim $\dzmax$? It is natural to consider the droplet deformation, which starts right upon impact, and therefore, could be the main cause for pushing grains away. If so, it would be plausible that $\dzmax$ scales with the kinetic energy imparted on the droplet $E_d = E_k - E_s = D_0/(D_0+2\zmax) E_k$ rather than that going into the sand. We find, however, that the data collapses nicely with $E_s$ and not at all with $E_d$ \footnote{In \cite{ZhaoSC2015}, it has been shown that this energy partition collapses the data of both the maximum droplet spreading diameter and the maximum crater depth, when packing fraction is varied. We suspect that the transient crater diameter data will only collapse with $E_d$ if the target's response is solid-like where solely the first layer(s) of grains are affected.}, see \fig~\ref{fig:maxZDD}\textit{d-e}, and hence, emphasizes the validity of our approach to divide the initial drop energy into deformation of both the sand bed and the droplet intruder. This result indicates that rather than droplet spreading it is the stress inside the granular bed that is redirected towards the free surface \cite{Murthy2012, Nordstrom2014, Clark2015} that determines the location of the transient rim. Furthermore, as the $\dzmax$ data smoothly collapses, the broadening of the final diameter $\dcinf$ versus $E_s$ that is observed for high energy in \fig~\ref{fig:dmax_final}\textit{b} is most likely due to the post crater formation evolution. 

\subsection{Crater slope \label{sec:slope}}
\begin{figure} \begin{center}
	\includegraphics[width=8.6cm]{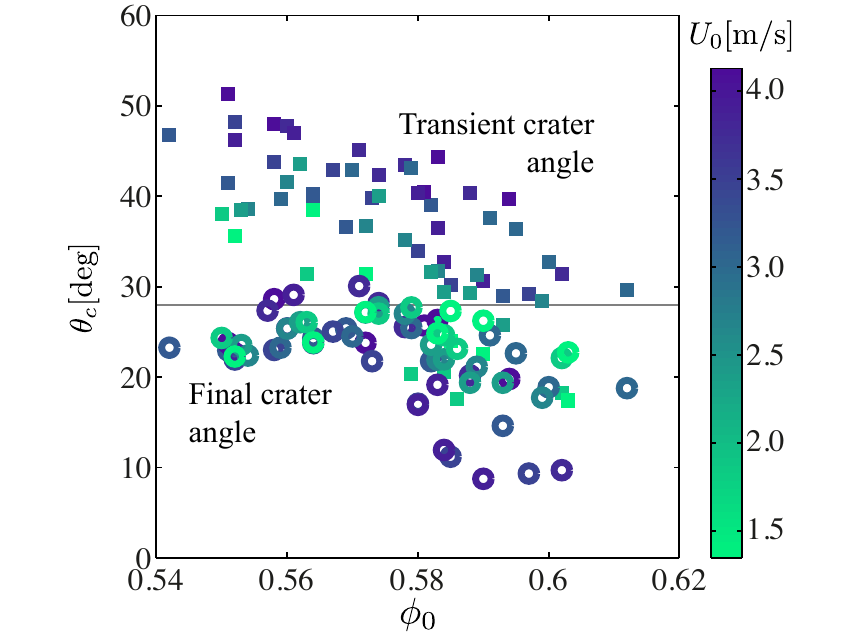}
	\caption{Final crater angle  $\cangle$ ({\scriptsize$\bigcirc$}) and the transient crater angle $\cangleD$ ({\scriptsize$\blacksquare$}) at the moment the maximum crater depth is reached, measured around the maximum crater slope, and plotted versus the packing fraction $\phi_0$. The first varies around a slightly smaller value than the angle of repose of $28\degree \pm 2\degree$ indicated by the horizontal solid line.
	\label{fig:c_angle} }
\end{center}
\end{figure}

The final diameter $\dcinf$ is always larger than the transient diameter $\dzmax$ (\fig~\ref{fig:maxZDD}). This implies that the crater broadens and that the slope of the crater walls changes from an initially steep to a flattened-off state, which can also be observed in \fig~\ref{fig:zofr_sd}. We extract the angle of each crater wall averaged around the maximum crater slope, thereby excluding the residue, and plot the result in \fig~\ref{fig:c_angle} as a function of packing fraction. The transient angle lies between $20\degree$ and $55\degree$, scatters with impact velocity and decreases for increasing packing density of the bed. The final angle varies around a value a bit below the angle of repose of $28\degree$ (which we measured using the funnel method), in contrast to the much smaller final crater angle reported in experiments of solid ball impact and the collapse of a cylindrical cavity in granular media~\cite{deVet2007, deVet2012}. It has been shown from a phenomenological model (of the BCRE-type \footnote{These BCRE type of models, named after Bouchaud, Cates, Ravi Prakash and Edwards, separate the granular substrate into a part that is moving down and a static phase.}) that the final crater slope depends significantly on the initial condition~\cite{Boutreux1997, Aradian2002}. Therefore, one plausible reason of our comparatively large final angle observations is that the transient crater angle never exceeds $50\degree$, while the collapse of a cavity created by solid ball impact are typically at a higher impact energy which may well cause much larger transient angles. Additionally, the spreading of the droplet may act as a stabilizing factor in the early crater formation, as upon impact a lot of its momentum is converted sideways.
In our results, the variation in the final angles is larger for dense packings ($\phi_0 > 0.58$), and the final angles are especially small for the largest impact speeds. We speculate that for those cases, the pancake-shaped residue plays a hard-to-observe-role in producing small final crater angles. In general, the fact that the final angles are close to the angle of repose indicates that avalanching acts as the main mechanism that produces the final crater shape. Combining this information with the observation that the transient depth is greatly affected by packing fraction, but the transient diameter not (as seen from \fig~\ref{fig:maxZDD}), it implies that avalanches (and maybe splashing) will modify the final crater shape more substantially for looser beds than for the denser ones. This can be seen from the difference between the transient slope and the final one. For example, it is plausible that for an initially deeper crater, more grains will avalanche than for a shallow transient crater (see also the next section). This might be the most important cause for the broadening of the data in \fig~\ref{fig:dmax_final}\textit{b}.

\subsection{Volume displacement \label{sec:volume}}
\begin{figure*}\begin{center}
	\includegraphics[width=1.0\textwidth]{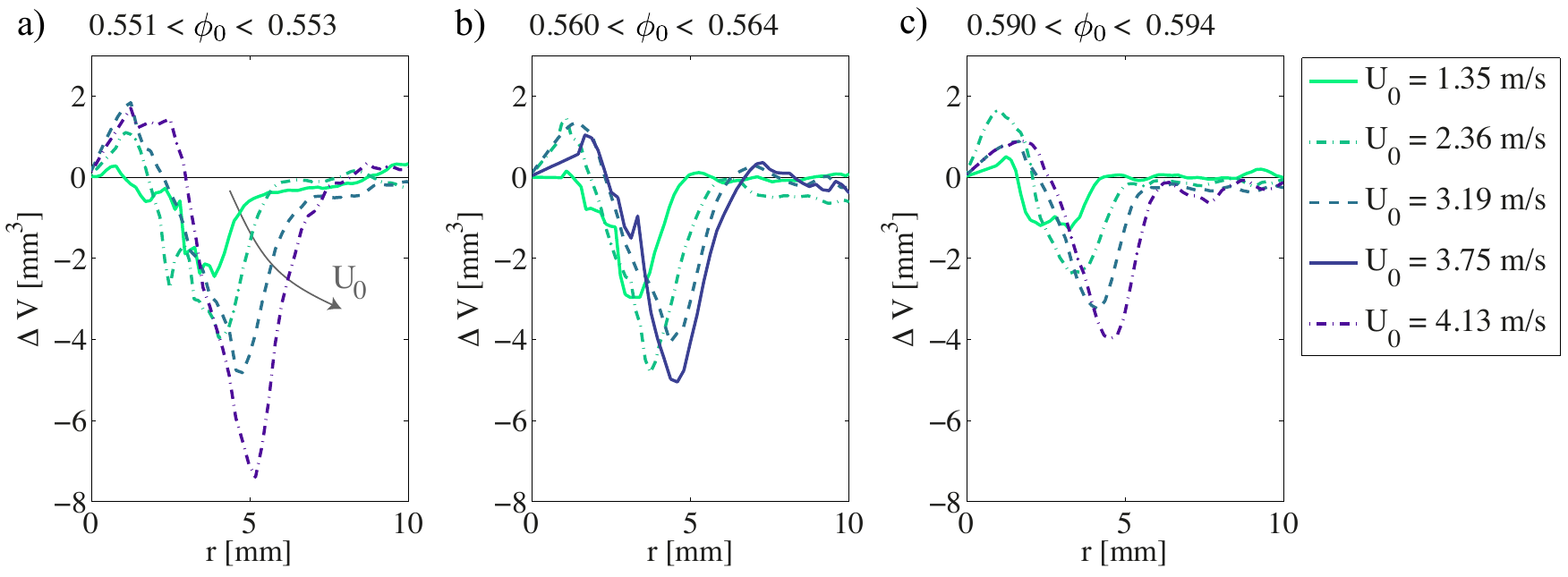} 
	\caption{The differential volume change $\Delta V$ between the final $\zst$ and transient $\zdy$ crater profiles are shown per radial distance for increasing impact velocity. It is denoted as: $\Delta V =2 \pi r \Delta r [\zst - \zdy]$ with $\Delta r = 0.18$~mm. The packing fraction is increasing from \abc{a} to \abc{c}.
	\label{fig:delVofr} }
\end{center}
\end{figure*}
Lastly, we want to discuss the volume displacement from the moment of reaching the maximum depth to the final crater shape. By comparing the transient and final profiles in \fig~\ref{fig:zofr_sd}, one can see that volume has moved away from the outer crater wall, \ie, near the rim, whereas volume is gained at the center. In \fig~\ref{fig:delVofr} we plot the differential volume between the final and transient shapes, $\Delta V = 2 \pi r \Delta r [\zst - \zdy]$, as a function of the radius \footnote{We choose to plot the radial volume difference and not the total volume, as minor systematic errors may cause a large variation in the total volume. E.g., for the final crater profile, the zero level has a small systematic uncertainty, which will however contribute significantly to the total crater volume when integrated over the radial coordinate.}. It shows clearly that the volume loss towards the rim is more significant than the central volume increase, and that the volume change is larger for looser packings, especially for higher impact velocities. The gained central volume in \fig.~\ref{fig:delVofr}\textit{a} also increases with impact velocity, which we believe originates from avalanches. We suspect that the packing fraction dependence is mainly caused by the residue, which contracts differently for truffle/doughnut and pancake shapes, namely sideways or a bit upwards respectively \cite{ZhaoSC2015}. 

Around the crater rim, a volume increase should be expected, both through grains that splash away and due to the compression at the center that causes grains to move outwards, towards the free surface. This volume is susceptible to errors (as it is relatively far from the center) and hence, cannot be determined reliably. However, as we observe that the volume loss is greater than the gain, one possible explanation would be that the volume is transferred outwards, to the side of the crater. It has to be noted that the total displaced crater volume is only expected to become zero for a sand bed at an initially critically packed state \ie\ $\phi_0 = \phi^*$, as shown by \cite{Umbanhowar2010, ZhaoSC2015}, as in this case the sand is likely to behave closest to an incompressible fluid.

Regardless the physical reason, we observe that at high impact velocities $U_0$, the crater evolution as it is expressed in, \eg, the slope change and volume loss, is much more pronounced and hence we can expect that the crater diameter data gets more dispersed for larger sand energies in \fig~\ref{fig:dmax_final}.

\section{Discussion \label{sec:discussion}} 
In addition to reporting and obtaining new insights by data acquisition, we also want to reflect and compare with previous work. We elaborate on two topics, the scaling of the final crater diameter with impact energy and the aspect ratio of the crater, in which case the above presented results may shine new light.

\subsection{Maximum crater diameter \label{sec:discusDc}}
In literature various scalings have been obtained for the maximum crater diameter after drop impact. Some report a scaling with Weber number ($\We = \rho_l D_0 U_0^2 / \sigma$) \cite{Katsuragi2011, Delon2011, Marston2010}, whereas others determine scalings with the initial kinetic energy of the droplet \cite{Nefzaoui2012, Zhao2015PNAS}. Since the impact velocity $U_0$ is the most prominently varied control parameter, we may look at the simpler relation of $\dcinf$ with velocity, \ie, $\dcinf \propto U_0^{\beta}$, where $\beta$ is found to range from about $0.34$ \cite{Zhao2015PNAS,Nefzaoui2012} up to $0.50$ \cite{Marston2010,Katsuragi2011, Delon2011}.
Also for variation with droplet size, different dependencies are reported. \citet{Nefzaoui2012} obtain an experimental fit of $\dcinf \propto E_k^{0.18}$ which matches most of their data, where, in the Cheng group, \citet{Zhao2015PNAS} find a fit of $D_c \propto D_0^{0.32} E_k^{0.17} / \left( \rho_l g \right)^{0.17}$, including an additional term containing the droplet diameter and $g$ the gravitational acceleration. In \cite{Katsuragi2011, Delon2011}, the drop size is simply included in the Weber number. 

The influence of the bulk density of the bed, $\rho_g \phi_0$, is studied by \citet{Katsuragi2011}, where in order to collapse the data a pre-factor in front of his $\We^{1/4}$ dependence is included, which consists of the granular bulk density over the liquid density. Packing fraction variation is considered only theoretically by \citet{Holsapple1987}, who expect that, for large energy impact of solid intruders on sand or rocks, an increasing compressibility should alter the scaling exponent from what they call energy to momentum scaling. 

In this work, we have varied impact velocity and packing fraction separately. For a given $E_k$, a large packing fraction dependence of $\dcinf$ can be observed in Fig.~\ref{fig:dmax_final}\textit{a} and \ref{fig:maxZDD}c, i.e., $\dcinf$ decreases with $\phi_0$, which is in contradiction to the bulk density dependence suggested by ~\citet{Katsuragi2010}. We find, however, that the data collapses with the part of the kinetic energy going into the sand, $E_s$ (cf. \fig~\ref{fig:dmax_final}\textit{b-c}), where the stiffness of the sand bed is taken into account. A power-law scaling of $\dcinf/D_0$ with $E_s^{\alpha}$ with $\alpha \approx 0.16 \pm 0.02$ is obtained. For the non-dimensional transient crater diameter a fit with $E_s$ gives a power-law exponent of $0.15 \pm 0.02$. \\

We believe that the variation in reported scaling laws may be related to the fact that in these works, it was not considered that not all initial kinetic energy $E_k$ of the impacting drop will be transferred into the sand, but also goes into drop deformation. The partition of $E_k$ will depend on the relative stiffness of intruder and target, where the one which is relatively easier to deform, will obtain most of the droplet kinetic energy. Here the extreme cases are the impact of a solid intruder on a granular bed and a deformable drop on a solid target. As we vary the packing density of our bed, this relative deformability is altered, since the stiffness of the bed increases at higher packing fractions. With the inclusion of the (experimentally measured) pre-factor in front of the kinetic energy, \ie, by introducing the portion of the impact energy that is transferred into the sand, $E_s =\frac{\zmax}{D_0/2 + \zmax} E_k$, we effectively account for the varying deformability that originates from changes in the packing fraction (\fig~\ref{fig:dmax_final}\textit{b}, \fig~\ref{fig:maxZDD}\textit{d}). 
Note that, $\zmax$ crucially depends on both kinetic energy and packing fraction, and with that also the just mentioned pre-factor $\zmax/(D_0/2 + \zmax)$. Therefore, the sand energy and initial kinetic energy of the drop do not have to be equivalent (\ie, $\dcinf$ does not scale with the same power with $E_k$ as with $E_s$). All in all, we think that the effect of the relative deformability together with avalanches, constitutes the origin for the variety in power laws that have been observed in previous studies for the maximum crater diameter. 

\subsection{Excavated mass}
In the dimensional analysis of crater formation by solid intruders, the impact energy is commonly balanced by either plastic deformation or by gravitational excavation \cite{Holsapple1993, Amato1998}, \ie, to lift grains against gravity. If the latter is the largest energy sink, and the crater dimensions would scale similarly, the dimensionless maximum diameter scales with the dimensionless energy with a power $1/4$ (as also observed in \cite{Walsh2003,Uehara2003}). In our case however, the power differs from $1/4$. Furthermore, we find that the excavated mass is not constant when packing fraction is altered: The crater volume decreases more with packing fraction than can be explained from the increase of the bulk density ($\rho_g \phi_0$). More specifically, as can be seen in Fig.~\ref{fig:maxZDD}, the transient crater volume, $\sim{\dzmax}^2\zmax$, decreases by about two thirds with an increase of $\phi_0$ of only 12\%. Therefore, $D_c^\infty$ will not collapse with the impact energy rescaled by the typical potential energy corresponding to the bulk density, $E_k/(\rho_g \phi_0 g {D_0}^4)$. Lastly, we can make a rough estimate of the gravitational potential energy of the final crater shapes, $E_p \propto \sum 2 \pi \rho_g \phi_0 g z^2_c(r,t^{\infty}) r \Delta r$, looking only at the excavated part (\ie, excluding the rim) and find that $E_p$ reaches at most $3\%$ of the initial impact energy. Moreover, $E_p$ decreases with packing fraction. These findings indicate that  dissipation processes during the deformation of the granular target are of paramount importance, and since the power-law exponent deviates from a $1/4$, the energy loss increases with $E_s$ in a non-linear manner.

\subsection{Crater depth versus diameter \label{sec:depthdiam}}
\begin{figure} \begin{center}
	\includegraphics[width=8.6cm]{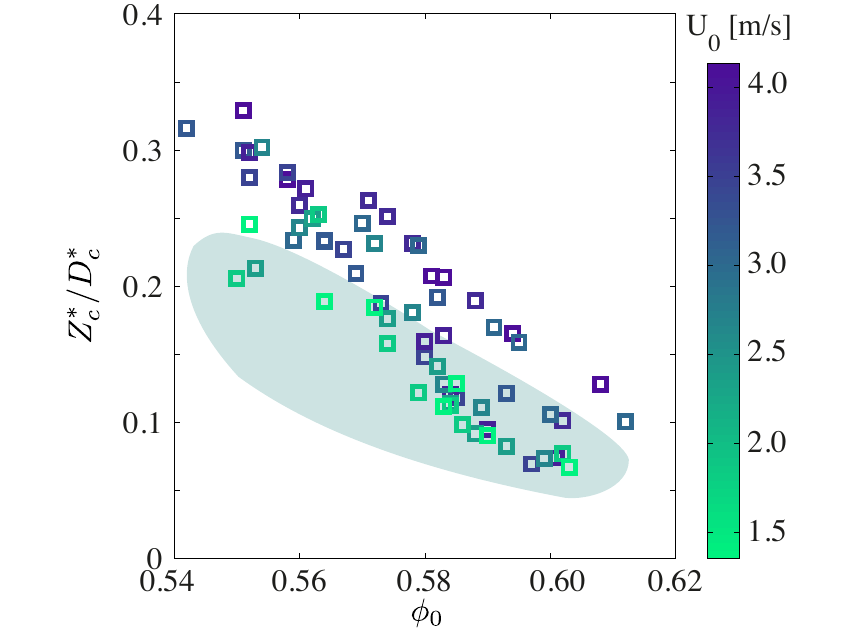}
	\caption{ The ratio between the maximum crater depth $\zmax$ and the crater diameter $\dzmax$ at the moment at which the maximum crater depth is reached, is plotted against packing fraction. The light (green) area shows the location of the data cloud for the ratio of $\zmax$ to the final crater diameter $\dcinf$, \ie, $\zmax / \dcinf$.
	\label{fig:zcdc} }
\end{center}
\end{figure}

For planetary and solid intruder impact craters, where the impact energy is usually much larger than for droplet impact, it has often been argued or assumed that the typical crater dimensions, \eg, diameter and depth, scale similarly with impact energy \cite{Amato1998, Holsapple1993, Schmidt1987}. There are studies that actually measure the final depth and diameter of the crater. From data of planetary craters, Refs \cite{Pike1980, Holsapple1993} reported a constant depth-to-diameter ratio for so-called simple craters. For smaller impact energies, namely for object impact onto sand, this ratio was only obtained for large grain size($>180\ \mu$m) \cite{Walsh2003}. However, \citet{deVet2007} find that depth and radius scale differently with energy. For the first they report an extra dependence on projectile density and for the latter with intruder radius. \citet{Uehara2003} investigated the maximum depth (measured below the intruder) and the final diameter and acquired different dependencies on impact energy. For droplet impact on sand, \citet{Zhao2015PNAS} reported again a constant ratio of depth to diameter of $0.2$, similar to \cite{Pike1980}, where they compare the depth of the crater at the moment the droplet/grain mixture rebounces (which happens for a subset of the impacts) and the final crater diameter. There they did vary the material properties of the sand, but the packing fraction was fixed to a high value.

In \fig~\ref{fig:zcdc}, we plot our result for the ratio of the maximum crater depth $\zmax$ and the diameter at maximum depth $\dzmax$ and observe that it has a clear dependence on packing fraction and scatters slightly with velocity. When comparing $\zmax$ with the final crater diameter $\dcinf$ as, \eg, was done in \cite{Zhao2015PNAS} (indicated by the green shaded area in \fig~\ref{fig:zcdc}), these dependencies do become smaller but do not disappear. The earlier mentioned ratio of $0.2$ lies within our data range.

To reconcile our findings with those of \citet{Zhao2015PNAS}, we compare the way in which the ratio has been measured. In  \cite{Zhao2015PNAS} the depth is acquired for the cases where a droplet/grain mixture bounces upward. This rebounce occurs at a much later time ($\sim 40$ ms after the impact) than the moment at which the maximum depth $\zmax$ is reached in our experiments (droplet spreading time scale $\sim 5$ ms). During this time interval the crater may be modified by \eg, avalanches and mixing of grains into the droplet. Therefore, these two depth measurements may differ in nature. However, as we did not observe droplet rebounce in our experiments, we can not directly compare these two different methods. To justify the results in \fig~\ref{fig:zcdc}, we need to clarify that there are two sources of uncertainties in the measurements of $\zmax$. One stems from the refraction of light through the liquid film which introduces a systematical underestimation of $\zmax$, and the other is introduced by the hyperbola extrapolation used for the unresolved crater centers (cf. \fig~\ref{fig:zofr_sd}). We estimated the uncertainties and find that the tendency shown in \fig~\ref{fig:zcdc} is robust~\footnote{The largest uncertainty introduced by refraction is $\sim 0.14$~mm which is very close to our depth resolution 0.1~mm. This results in a slight underestimation of the maximum depth that is proportional to the thickness of the liquid film at maximum spreading which decreases with packing fraction and impact energy. For a more detailed discussion, we refer to Ref. 25 in \cite{ZhaoSC2015}. Note that the reflectance from air to water is negligible, as according to the Fresnel' equations, it is $<10\%$ and in our experiments the deformed liquid is concave, resulting in an even smaller effect. To estimate the uncertainty of extrapolation, we select a case where the crater center depth is well measured. We fit a hyperbola to the crater profile for $z,0$ and $r>1$ mm. The difference between the fitted depth and the measured one is 0.06 mm. Both uncertainties are less than 2\% of $\dzmax$ and $\dcinf$ and hence, their influence on the aspect ratio is negligible.}. We want to stress that \citet{Zhao2015PNAS} obtained the depth only for a subset of their experiments, where we acquired data for all impacts that we have studied. 

To understand the reason for a non-constant aspect ratio in our data, one may refer to the opening phase of the crater. We see that the transient crater diameter, $\dzmax$, collapses simply with $E_s$, and no packing fraction dependence is observed (\fig~\ref{fig:maxZDD}\textit{d}). For the maximum depth $\zmax$, however, we observed that there is both a (distinguishable) dependence on sand energy and on packing fraction (see \fig~$5$\textit{b} in \cite{ZhaoSC2015}). This suggests that the crater depth and diameter are determined by different mechanisms, at least for low intruder energies. For high energy or planetary impacts, the material around the impactor might rather get destroyed or molten and hence, the liquid-like material below and beside the impactor would respond similarly leading to the observation of a more constant aspect ratio.

\section{Conclusion}
In this work, we have shown that in crater formation after the impact of a droplet onto a granular material subtle and complex processes play a role. We compared the transient crater shape, determined at the moment that the crater reaches its maximum depth, and the final crater profile. By including a measure for the relative stiffness of intruder and target, we could approximate the portion of the  initial kinetic energy going into sand bed deformation. As can be seen in Figs \ref{fig:dmax_final} and \ref{fig:maxZDD}\textit{d-e}, both the transient and final crater diameter collapse with this quantity. For the transient crater diameter all packing fraction dependence is captured by using the energy division, however, for the final diameter at the largest sand energies a dependence is still observed. Avalanching is likely to be responsible for this diameter evolution, as it becomes more pronounced for looser beds (\fig~\ref{fig:c_angle}). In addition, we surprisingly observed that in the transient phase, the crater may respond differently in the downward and outward direction, as in \fig~\ref{fig:maxZDD}\textit{a-b} the depth displays a clear packing fraction dependence which is not seen for the diameter.

This work is financed by the Netherlands Organisation for Scientific Research (NWO) through a VIDI Grant No. 68047512.


%

\end{document}